\documentclass[12pt]{article}
\usepackage{cite}
\usepackage{amsmath,amssymb,amsfonts}
\usepackage{algorithmic}
\usepackage{graphicx}
\usepackage{textcomp}
\usepackage{xcolor}
\usepackage[utf8]{inputenc}
\usepackage{hyperref}
\usepackage{authblk}
\usepackage[margin=1in]{geometry}
\usepackage{enumitem}
\usepackage{xcolor}
\usepackage{fancyhdr}
\begin{document}

\chead[C]{\scriptsize \textcolor{red}{\textit{Draft under review in the \textbf{Biomedical Signal Processing and Control} journal. DO NOT REDISTRIBUTE}}}
\fancyfoot[C]{\quad\quad\quad\quad\quad\quad\thepage}
\renewcommand{\headrulewidth}{0pt}
\renewcommand{\footrulewidth}{0pt}
\renewcommand{\sectionmark}[1]{}
\renewcommand{\subsectionmark}[1]{}

\title{\textbf{Comparing sleep studies in terms of the Apnea-Hypopnea Index using dedicated Shiny web application}}

\author[$1,\star$]{Marcel Młyńczak}
\author[$2$]{Tulio A. Valdez}
\author[$3$]{Wojciech Kukwa}

\affil[$1$]{\small{Warsaw University of Technology, Faculty of Mechatronics, Institute of Metrology and Biomedical Engineering, 8 Boboli Street, 02-525 Warsaw, Poland, \texttt{marcel.mlynczak@pw.edu.pl}}}
\affil[$2$]{\small{Stanford University, Department of Otolaryngology, Palo Alto, California, US}}
\affil[$3$]{\small{Medical University of Warsaw, Department of Otorhinolaryngology, Warsaw, Poland}}

\date{}
\maketitle

\begin{abstract}
The Apnea-Hypopnea Index (AHI) is one of the most-used parameters from the sleep study that allows assessing both the severity of obstructive sleep apnea and the reliability of new devices and methods. However, in many cases, it is compared with a~reference only via a correlation coefficient, or this value is at least the most emphasized. In this paper, we discuss the limitations of such an approach and list several alternative quantitative and qualitative techniques, along with their interpretations. We propose the assessment of clinical significance along with the statistical one. Qualitative analysis can be used for this purpose, or we suggest using the ranking function which enables consideration of various AHI values with different weights. It can be reliable for both adult-related and pediatric sleep studies. The dedicated Shiny web application, written in R, was developed to enable quick analysis for both physicians and statisticians. \\ 

\noindent \textbf{Keywords}

\noindent Apnea-Hypopnea Index, correlation coefficient, Bland-Altman analysis, clinical significance, ranking function, Shiny web application   \\

\noindent \textbf{Highlights}

\begin{itemize}[leftmargin=*]
	\item AHI is the most popular parameter assessing the sleep-disordered breathing severity.
	\item It is used to evaluate the new devices, but too often by only a correlation coefficient.
	\item Other context-increasing qualitative and quantitative approaches are discussed.
	\item Shiny Web application is provided to enable quick calculations and comparisons.
	\item The clinical significance may be tested, e.g., by applying the ranking function.
\end{itemize}

\end{abstract}

\section{Introduction}

Obstructive sleep apnea (OSA) is the most severe form of sleep-disordered breathing. It is a common health problem, moderate to severe forms affecting $49.7\%$ of men and $23.4\%$ of women, according to the latest epidemiological data \cite{problem}. This condition affects all age groups, with a prevalence of $2-4\%$ in children, and a peak prevalence occurring at 2-8 years of age for pediatric patients \cite{ped1,ped2,ped3,ped4,ped5}.

The gold standard study for diagnosing OSA is polysomnography (PSG), and one of the main parameters is the Apnea-Hypopnea Index (AHI). It allows assessment of the degree of sleep-disordered breathing, particularly considering the number of apneas and hypopneas within a single “average” hour during sleep. It is generally accepted and widely used. Based on the medical guidelines, there are four qualitative subranges of sleep apnea for adults \cite{AHI}:

\begin{itemize}
	\item Normal: $AHI < 5$,
	\item Mild: $5 \leqslant AHI < 15$,
	\item Moderate: $15 \leqslant AHI < 30$, and
	\item Severe: $AHI \geqslant 30$.
\end{itemize} 

\noindent Different thresholds are used for the same classifications in the pediatric population:

\begin{itemize}
	\item Normal: $AHI < 1$,
	\item Mild: $1 \leqslant AHI < 5$,
	\item Moderate: $5 \leqslant AHI < 10$, and
	\item Severe: $AHI \geqslant 10$.
\end{itemize}

PSG is a very complex and costly test. Therefore, many different abbreviated sleep studies (usually home sleep tests, HST) are being developed, and all of these need to be validated against full polysomnography. Therefore, new devices and/or methods evaluating obstructive sleep apnea (and sleep disordered breathing in general) are undergoing comparative studies wherein they are used to measure AHI, and assessed against reference equipment.

\subsection{Correlation coefficient}

However, despite the guidelines, still the exact (raw) values are compared, very often using correlation coefficients, like in a recent meta-analysis of peripheral arterial tonometry diagnostics \cite{Wrong} or an assessment of portable wireless sleep monitors \cite{odWojtka}. 

There is nothing inherently wrong with the analysis of numbers instead of diagnostics subranges; however, it appears that the meaning of the correlation coefficient was overlooked in those cases. Several issues should be addressed in the context of AHI comparison. Firstly, correlation coefficient does not evaluate the clinical significance of the method, e.g., how often subjects are correctly classified (as normal, mild, moderate, or severe). It is also susceptible to outliers and influential observations. Pearson’s version should be used only for data with a normal distribution, a condition usually unfulfilled due to a limited range of physiologically relevant values and the spectrum of the invited study group. Also, its interpretation is connected with a linear model - many regression lines, with different slopes and intercepts, may produce high correlation coefficients, even if not clinically relevant (slope of 45 degrees without the intercept). Additionally, there is no insight into whether the slope is slightly greater than or (to the same extent) less than 1, which might otherwise drive important conclusions on the reliability of the tested device.

\subsection{Qualitative approaches}

Therefore, we asked the question - how to compare results from sleep studies, and to emphasize possible clinical (not only statistical) errors? 

Parameters of the possible quantitative approaches that still work with raw values, but do not (or to a much lesser extent) introduce the aforementioned misinterpretations, are (the order is arbitrary): the intercept of the linear model that best fits data points; the slope of another linear model with the intercept value forced to 0; the p-value (or test statistics) from a Wilcoxon rank paired test (or from a paired T-test in case of normally distributed data); the rho value (along with its p-value) from Spearman's rank correlation test (not assuming a normal distribution); Lin's Concordance Correlation Coefficient (with its confidence interval limits), measuring how far the data deviate from the line of perfect concordance (line at 45 degrees) \cite{Lin1,Lin2}; the bias correction factor (from Lin's analysis), that determines how far the best-fit line deviates from a line at 45 degrees (no deviation from the 45 degree line occurs when factor equals 1) \cite{Lin1,Lin2,DescTools}.

Also, the following quantitative ones might be mentioned: mean difference of AHI scores and $\pm1.96SD$ of AHI score differences on a Bland-Altman plot (means of AHI scores on the X-axis, and differences of AHI scores on the Y-axis) \cite{BA}; slope and intercept of the linear model that best fits data points on the modified Bland-Altman plot (in which reference values are used instead of the mean AHI scores on the X-axis) - this technique enables one to assess whether the nature of the differences' distribution depends strictly on the reference value; simple heuristic ratio - the number of data points above the $Y = X$ line divided by the number of points below it; or mean absolute error (MAE). Another technique, the relative-deviation Bland-Altman plot, may also be used to complete the analysis with regards to the visual distribution of points \cite{reldevBA}.

\subsection{Quantitative approaches - clinical significance}

\begin{figure}[!h]
\centerline{\includegraphics[width=11.9cm,height=9.4cm]{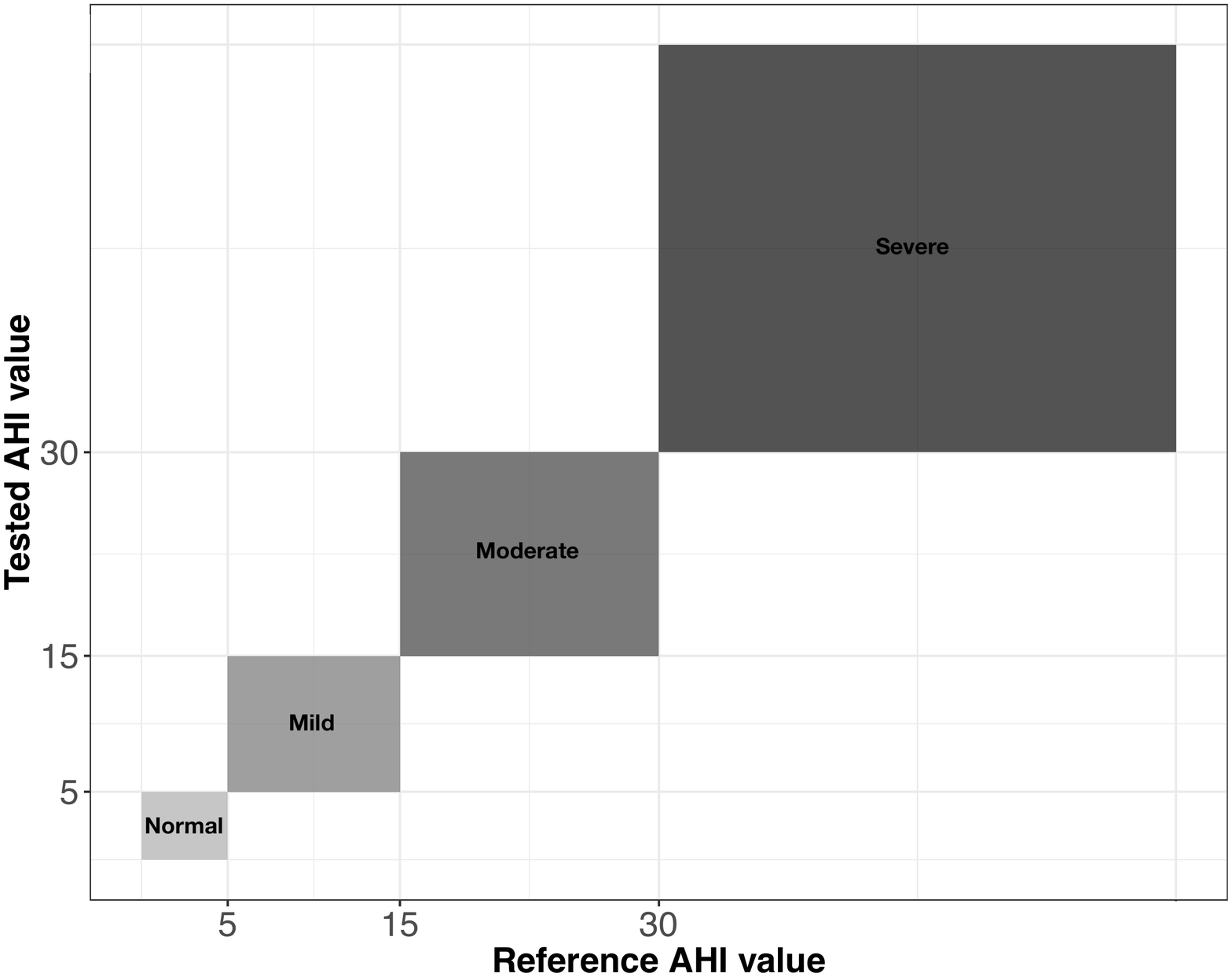}}
\caption{While assessing the clinical significance of the results compared to the reference, one needs to check whether the points are inside presented squares, even if the numerical differences are relatively large (but still within the same clinical interpretation - AHI subrange). The figure presents squares for adults.}
\label{fig1}
\end{figure}

While the techniques listed above are mathematically correct, all of them use the numerical data directly, without supplying the right context (established subranges of AHI and their clinical interpretations). Therefore, qualitative techniques are also important, being able to estimate “clinical significance” along with “statistical significance”. The introductory concept is presented in Fig. \ref{fig1}. It is an illustration of squares, that if the values from both measurements are inside, the same clinical approach to a patient would be performed (“same clinical approach” means that in the medical guidelines the same medical procedure is proposed when the subjects are within the same AHI subranges - it doesn't matter if one has, for example, AHI 16 and the other 28). The main idea behind the figure is that the method can “clinically” comparable to the reference when points are within squares, not only when as close to Y = X line as possible.

The main parameters - accuracy, sensitivity, specificity, positive and negative predictive values (PPV and NPV, respectively), or Cohen's Kappa - may then be estimated for the specific case. The accuracy is the ratio of the number of correctly classified data points to the total number of data points. Even if there is a relatively big difference, as between 16 and 28, the assignment to a group will be the same, unlike for 14 and 16, which could result in the patient receiving different treatment. The sensitivity can be calculated as the proportion of true positives; it can be extended to a multi-variables case by making the positive state the one being analyzed, and grouping all others into the negative. The extended specificity uses the same definition, with true positives replaced by true negatives.  

PPV and NPV are the proportions of predicted positive and negative results in the test to the true positive and true negative results, respectively. Cohen’s Kappa is a more robust value, which describes the accuracy after removing the effect of random choice and takes into account possible imbalances in the data; all values greater than 0 mean that the method is better than a coin toss (the maximum is 1, the same as for accuracy).

Additionally, multi-class Receiver Operating Characteristic (multi-class ROC) analysis may provide the parameter of area under the curve (AUC), along with pair-wise ROC curves \cite{multi,pROC}.

\subsection{Objectives}

This work aims to discuss approaches of the AHI comparison, and their mathematical and clinical interpretations; to propose a novel ranking function to enhance values around different clinical managements; and to present the Shiny web application, which contributes to the field by implementing selected methods and enabling extended analysis of data.

\section{Materials and Methods}

\subsection{Shiny web application}

The dedicated Shiny web application \cite{Shiny} (written in R with external packages: \textit{shiny} \cite{pshiny}, \textit{shinythemes} \cite{pshinythemes}, \textit{ggplot2} \cite{pggplot2}, \textit{plotly} \cite{pplotly}, \textit{DT} \cite{pDT}, \textit{BlandAltmanLeh} \cite{pBlandAltmanLeh}, \textit{caret} \cite{pcaret}, \textit{e1071} \cite{pe1071}, \textit{DescTools} \cite{DescTools}, \textit{pROC} \cite{pROC}, \textit{grid} \cite{pgrid} and \textit{ggExtra} \cite{pggExtra}) was developed primarily for doctors and clinicians in order to enable work with these methods using only two vectors - reference and measurement AHI values. 

\begin{figure}[!h]
\centerline{\includegraphics[width=15.52cm, height=9.13cm]{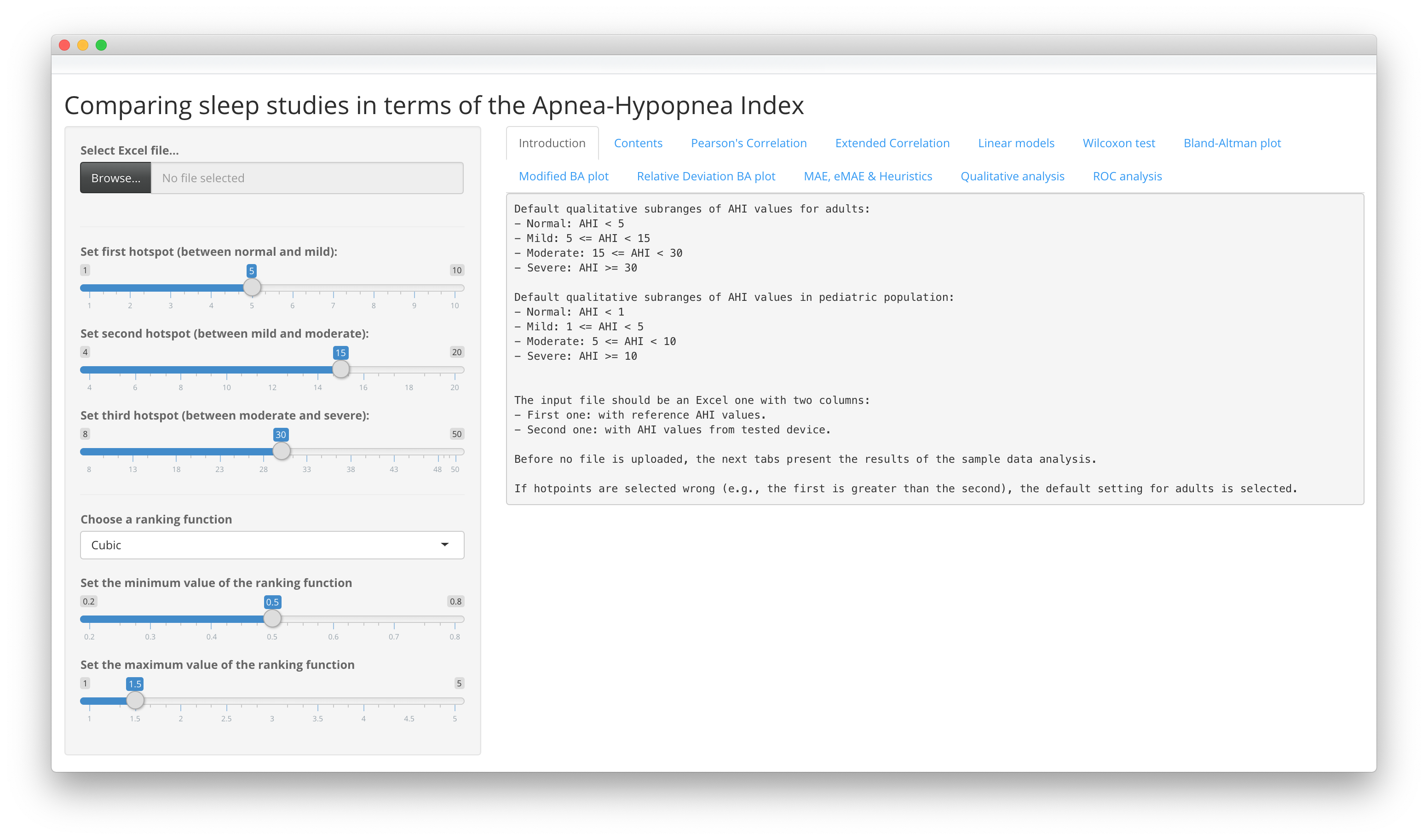}}
\caption{The home page of the Shiny web application, supplemental for the paper \cite{Shiny}.}
\label{shiny}
\end{figure}

The home page, containing inputs (left panel), tabs (upper right) and the output display (lower right) is shown in Fig. \ref{shiny}.

The input file should be an Excel spreadsheet with two columns comprising reference AHI values and values from a tested device. Before any file is uploaded, the tabs present the results of the analysis of the default data, coming from \cite{Data}, a study assessing the agreement between the AHI parameters measured by portable monitor and by reference polysomnogram. The data are not well concordant, which facilitates presentation.

In the left panel, the app enables choosing all thresholds to determine which subrange the specific result is located in (by default, these are $5$, $15$ and $30$, as presented in the Introduction for the adult-related group). It is also possible to set the minimum and maximum of the ranking function (by default, $0.5$ and $1.5$, respectively), and the approximation approach (the function is described in the next section).

\textbf{Contents} provides the preview of the Data. \textbf{Pearson's Correlation} tab can be used for estimating the coefficient and the p-value of the statistical test. \textbf{Extended Correlation} tab allows estimation of Spearman's correlation coefficient (rho) and p-value, along with Lin's coefficient, confidence intervals, and bias correction factor (which measures how far the best-fit line deviates from a line at 45 degrees. No deviation from the 45 degree line occurs when factor equals 1). \textbf{Linear models} calculates intercept of the linear model when intercept is included, slope of another model when intercept is excluded; and both equations with the figure are provided. \textbf{Wilcoxon test} tab implements Wilcoxon rank paired test p-value calculation (with interpretation), and the histogram of differences is also presented.

\textbf{Bland-Altman plot} presents the results of the method. \textbf{Modified BA plot} also presents the results, but in comparison to the original method, reference AHI values create X axis and slope of the linear model is calculated. \textbf{Relative Deviation BA plot} enables presentation of relative differences - this is related with Bland-Altman plot. \textbf{MAE, eMAE \& Heuristics} tab collects mean absolute error, extended mean absolute error (after applying ranking function described in the next section) and heuristic ratio - being a ratio between values above and below Y = X line. \textbf{Qualitative analysis} presents several results of the analysis based on defined subranges of AHI values, e.g., accuracy, Cohen's Kappa, statistics by class etc. \textbf{ROC analysis} allows calculating multiclass-ROC area under the curve, supplemented by pair-wise ROC figures.

Other visual/analytical methods are possible (like finding the distribution of absolute or square differences, Bayesian reasoning, etc.); due to their poorer interpretability from the physician's perspective, they were not considered further.

\subsection{Ranking function}

Coming back to the statement that the difference between 16 and 28 may be less significant from the clinical point of view than that between 14 and 16 (which may cause different treatment to be prescribed), and considering an approach based on numerical values, we would like to propose a ranking function that will allow the introduction of weights, which should be multiplied by the original difference, increasing its impact around hotspots (for standard adult-related AHI analysis, they are 5, 15 and 30) and decreasing impact in the middles of the subranges.

\begin{figure}[!h]
\centerline{\includegraphics[width=10.8cm,height=7.2cm]{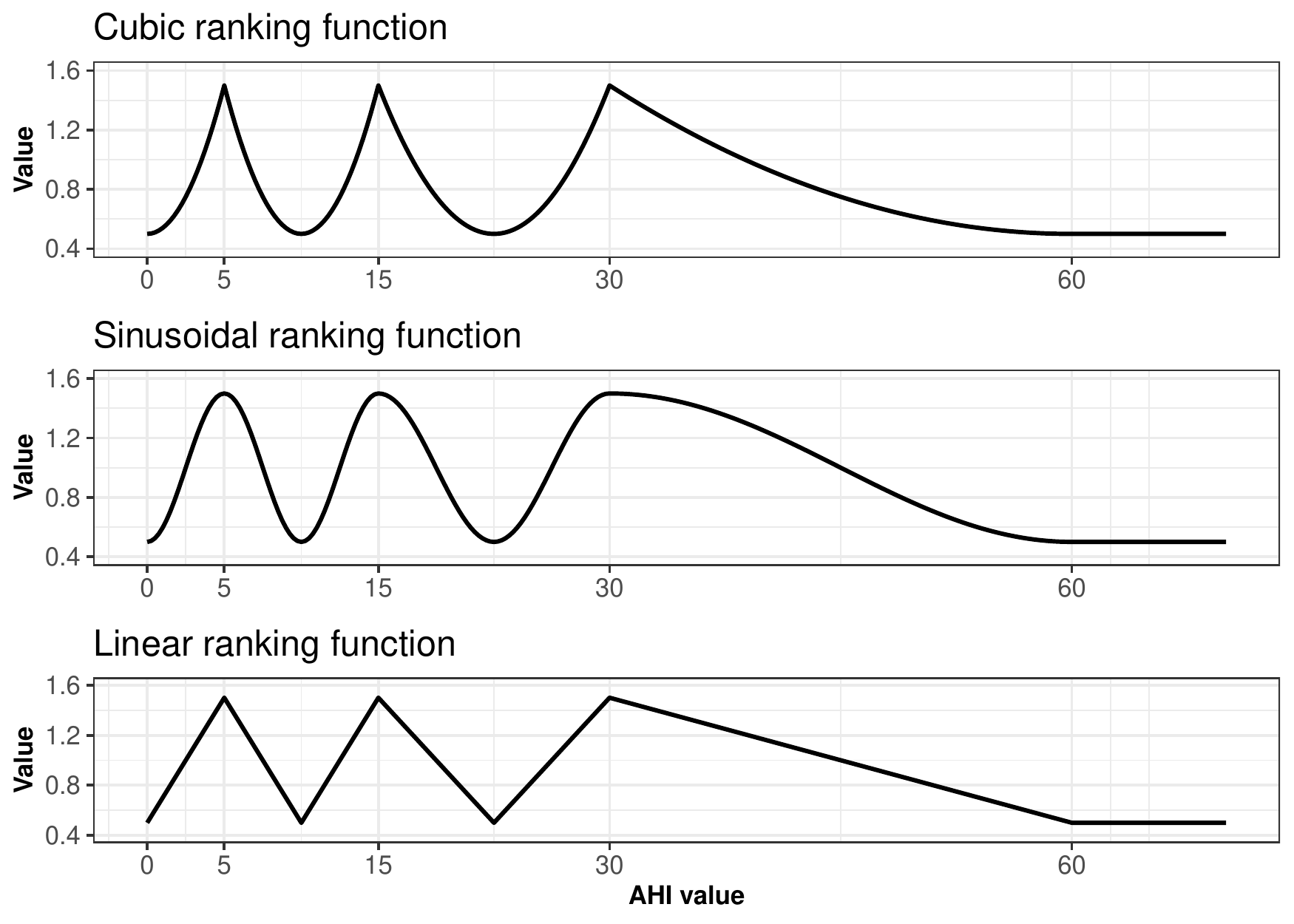}}
\caption{The shapes of the proposed ranking functions - three different approximations - cubic (default), sinusoidal and linear; the X-axis value is the reference AHI and the Y-axis presents the coefficient for multiplying the difference.}
\label{fig2}
\end{figure}

Of course, the definition of such a ranking function may vary. There may be cubic, sinusoidal or linear approximations between hotspots. All are available to be selected in the application. We decided to set a default value of $1.5$ at hotspots and a value of $0.5$ at range midpoints. We also decided, that the function would equal $0.5$ after twice as big as the third hotspot - 60 by default for adult population. The shapes of such functions are presented in Fig. \ref{fig2}. The graph is produced by: (1) taking into account hotspots (for adults 5, 15 and 30 by default); (2) establishing extrema values; (3) interpolating the function in between based on the chosen method (cubic by default). For instance, when finding the formula for the range between AHI 5 and 15, the parabola is searched to cross 3 points (5 and 15 at maximum, and 10 (middle) at minimum. 

Further generalization can take into account the case in which each of the points is near a different hotspot, with both still in the same subrange. For clarity and simplicity, we propose to apply the multiplier of $0.5$, as explained below, for the extended mean absolute error (eMAE) formula \eqref{Eq1}  

\begin{equation}
eMAE\, =\, \frac { 1 }{ n } \sum _{ i=1 }^{ n }{ A\left( { e }_{ i } \right) \cdot B\left( { ref }_{ i } \right) \cdot \left| { res }_{ i }-{ ref }_{ i } \right|  } \label{Eq1}
\end{equation}

\noindent where $n$ is the number of participants in the study, $res$ is the measured AHI value, $ref$ is the reference value, $i$ is the iterator counting successive participants, $B$ is the ranking function, and $A$ is the function with a value of $0.5$ for points ($e$) in the same subrange of AHI values, and $1.0$ for others. Of course, the chosen values may be different, but we decided to set them arbitrarily.

\subsection{Testing procedure}

The motivation of the testing procedure is to show how to quantify the performance of methods for automatically estimating the AHI. The elaborated solution was carried out on two sample independent (in terms of the method and subjects) datasets: (1) on the data from the paper \cite{Data} (available online), and (2) from \cite{Nakano} (received from authors). The former consists of 71 observations, the latter - 304 points coming from testing subset. The demographics of subjects included in the studies are reported in Table \ref{tab-demo}.

\begin{table}[h]
\centering
\caption{The summary of demographic parameters of the participants included in the assessed studies \cite{Data,Nakano}}.
\label{tab-demo}
  \begin{tabular}{l|cc}
    & \textbf{Dataset 1} \cite{Data} & \textbf{Dataset 2} \cite{Nakano} \\
    \hline
  N  		   						& 71 & 304 \\
  Age [Y; $mean \pm SD$]  			& $52 \pm 10$& $54.8 \pm 15.4$\\
  Sex [F/M]  						& 29 / 42 & 64 / 240 \\
  Normal [N]  						& 6 & 32\\
  Mild [N]  						& 17 & 60\\
  Moderate [N]  					& 9 & 62\\
  Severe [N]  						& 39 & 150\\ \hline
  \end{tabular}
\end{table}

The Results section presents the most important results produced for two various sample datasets (and the eMAE is compared for all considered approximation approaches), using the app \cite{Shiny}. The reason why we chose the two datasets were to present worse and better results, which came from two independent methods.

\section{Results}

\begin{figure}[!h]
\centerline{\includegraphics[width=0.9\columnwidth]{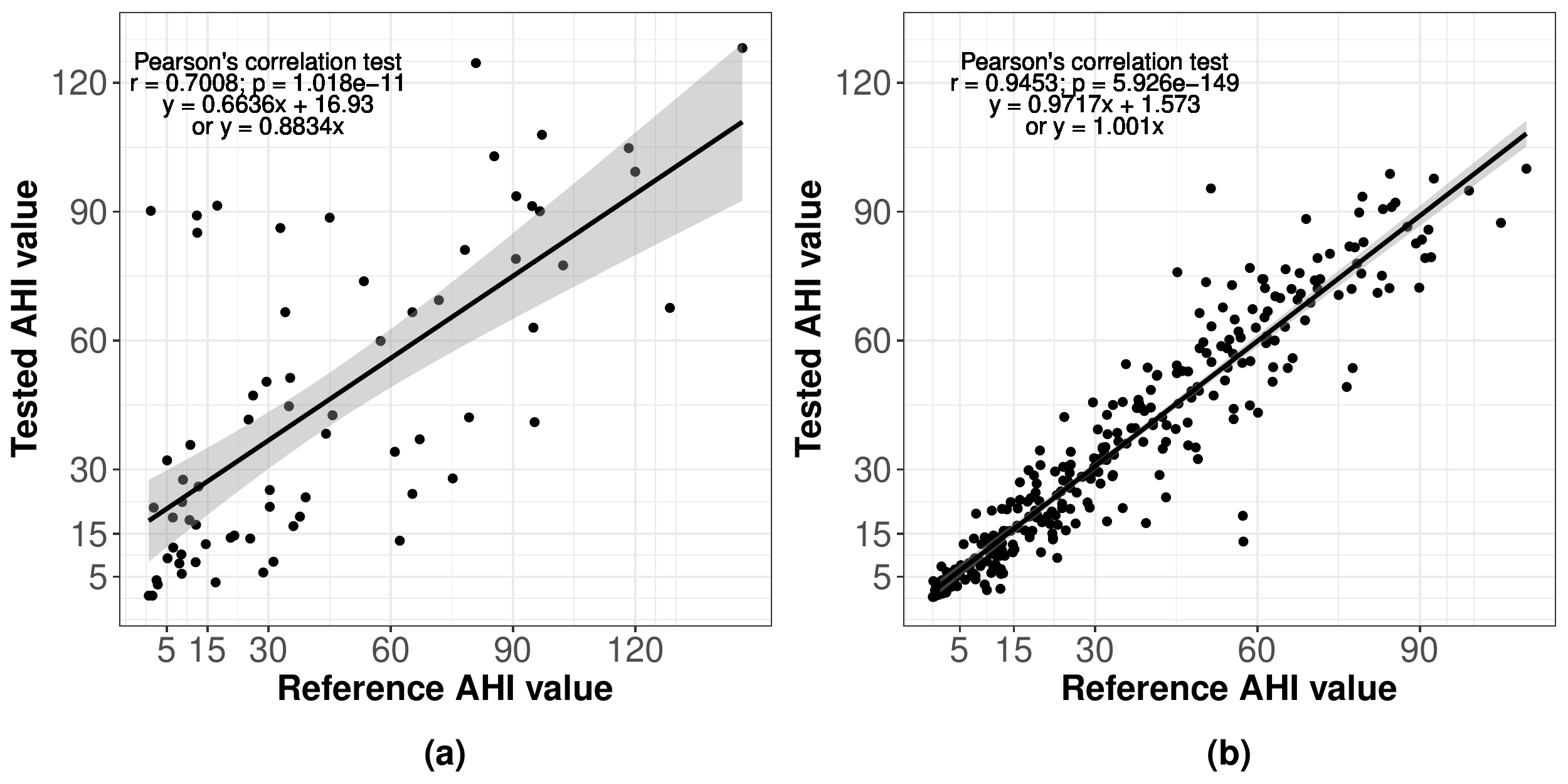}}
\caption{The scatter-plots and basic statistics of the data points (in the Shiny web app, the plot is divided into two tabs: Pearson's Correlation and Linear models). a) On the left - the analysis on the first dataset; b) on the right - the analysis on the second dataset.}
\label{fig3}
\end{figure}

Scatter-plots of data are presented in Fig. \ref{fig3}, along with the Pearson’s correlation coefficients, the p-values of the correlation test, the formulas and lines for the simple linear regressions with intercept, and only the formulas for another simple linear regressions calculated without considering intercept.

One can note that for the former dataset the intercept of the first considered linear model was $16.928$, even higher than the second hotspot. On the other hand, the slope of the model with the intercept value forced to 0 is $0.883$. The latter dataset has close slopes, $0.97$, and $1.001$, for the model with and without considering intercept in the linear model, respectively.

The Spearman's rho were about $0.68$, and $0.95$, respectively. In both cases, it was similar to the Lin's coefficient, which does not differ much from the other correlation measures, as the bias correction factor is very close to 1, regardless of poor data concordance for the first dataset.

As the data did not come from a normal distribution, a Wilcoxon rank paired test should be performed. The p-values of $0.95$ and $0.04$ were reported, respectively. It indicates that for the first case there is no cause to reject the null hypothesis - exact medians; however, slight, but statistically significant difference is observed - probably due to higher N.

\begin{figure}[!h]
\centerline{\includegraphics[width=0.8\columnwidth]{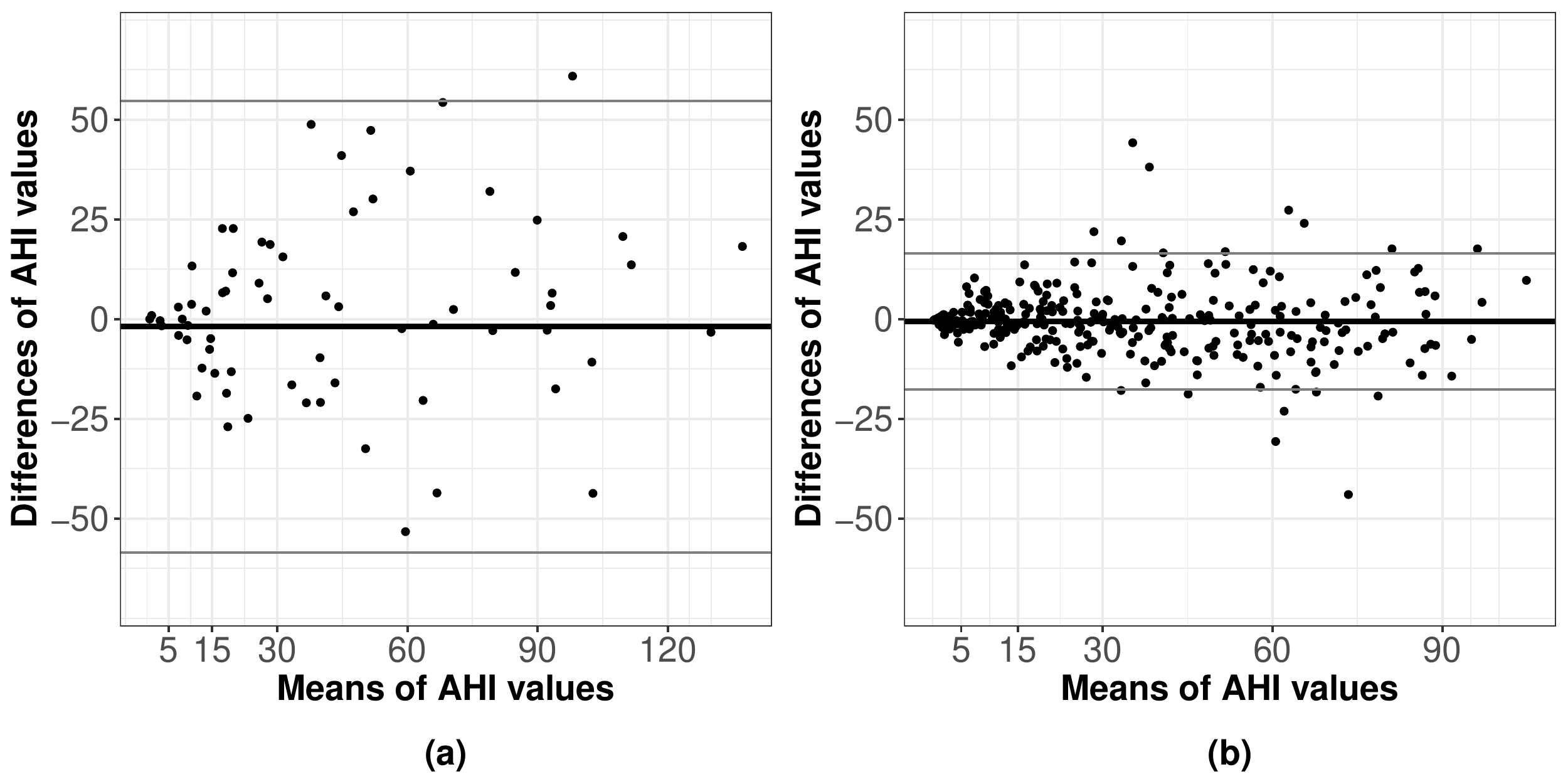}}
\caption{The Bland-Altman plots of the analyzed data points. a) On the left - the analysis on the first dataset; b) on the right - the analysis on the second dataset.}
\label{fig4}
\end{figure}

The Bland-Altman plots are presented in Fig. \ref{fig4}. The mean value of the differences (without using their absolute values) are $-1.91$ and $-0.59$, respectively; and the mean spread calculated by the $\pm1.96SD$ measure are $56.62$ and $17.03$, respectively. For the first dataset it is very high, which can also be observed in the Bland-Altman plot's high percentages of relative deviation.

\begin{figure}[!h]
\centerline{\includegraphics[width=0.8\columnwidth]{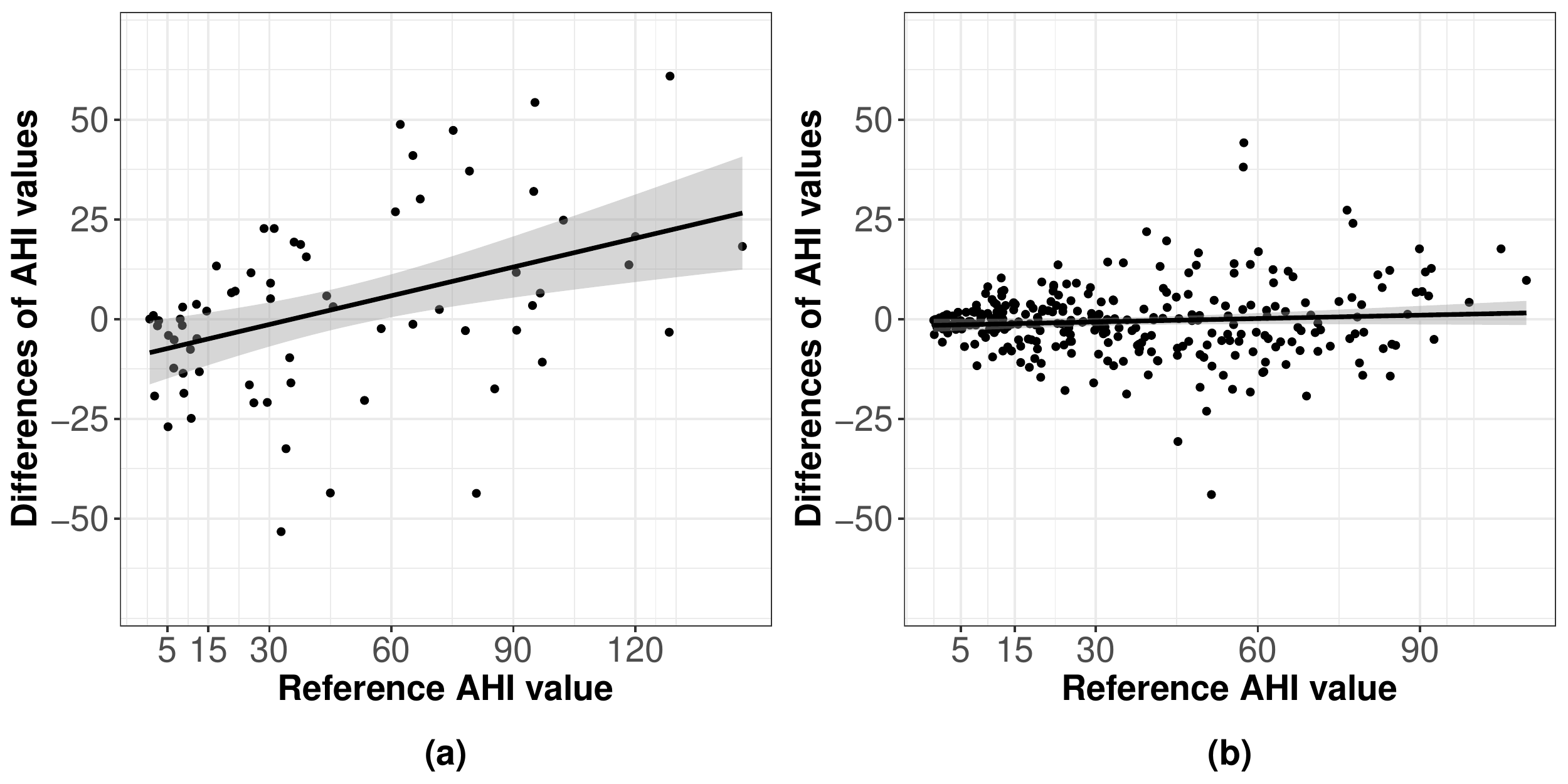}}
\caption{The modified Bland-Altman plots of the analyzed data points, along with the linear model. a) On the left - the analysis on the first dataset; b) on the right - the analysis on the second dataset.}
\label{fig5}
\end{figure}

For the modified Bland-Altman (Fig. \ref{fig5}), the slope of the linear model for the first considered dataset was relatively high, about $0.34$, showing that the difference between values is related to the actual value. It does not apply for the latter dataset (which is positive).

The values of MAE and eMAE for all considered ranking functions are gathered in Table \ref{tab-mae}.

\begin{table}[!h]
\centering
\caption{The values of mean absolute errors and extended mean absolute errors (for all ranking functions - cubic, sinusoidal and linear). The relative differences are estimated in relation to Dataset 2.}
\label{tab-mae}
  \begin{tabular}{l|ccc}
    & \textbf{Dataset 1} & \textbf{Dataset 2} & \textbf{Relative difference} \\
    \hline
  \textbf{MAE}  		   		& $20.21$ & $5.91$ & \textit{2.42x} \\
  \textbf{eMAE} \textit{(Cubic)}  	& $11.90$ & $2.76$ & \textit{3.31x} \\
  \textbf{eMAE} \textit{(Sinusoidal)} & $13.88$ & $3.24$ & \textit{3.28x} \\ 
  \textbf{eMAE} \textit{(Linear) }	& $13.85$ & $3.23$ & \textit{3.29x} \\ \hline
  \end{tabular}
\end{table}

Presented results show that eMAE enable stressing the errors more than regular MAE, and cubic ranking function little more than sinusoidal and linear, as the mean value of ranking function throughout the range is the lowest (the differences are the highest around hotspots).

For qualitative analysis, the three main parameters for both datasets were as stored in Table \ref{tab-res}.

\begin{table}[!h]

\centering
\caption{The summary of qualitative analysis of both considered datasets, and assuming default boundaries of subranges as for adults.}
\label{tab-res}
  \begin{tabular}{l|cc}
    & \textbf{Dataset 1} & \textbf{Dataset 2} \\
    \hline
  Accuracy [$\%$]  		   	& $57.7$ & $84.2$ \\
  Cohen's Kappa  			& $0.32$ & $0.76$ \\
  Multi-class AUC  			& $0.733$ & $0.939$ \\ \hline
  \end{tabular}
\end{table}

It shows, that even though several methods prefer to treat the differences as statistically insignificant (like for the Wilcoxon test as for the first dataset), the assessment of clinical significance (e.g., difference between accuracy and $100\%$) provides noticeable results - for the first dataset almost half of points would have different diagnosis, and for another - even if the quantitative measures are very promising, the presence of approximately $15\%$ of bad interpretations seems to be indisputable.

\section{Discussion}

The PSG is still a gold standard in sleep research, even if it is too complex for breathing-related studies - the final analysis being based on several parameters, e.g., the apnea-hypopnea index, respiratory disturbance index, or percentage of snoring during the night, from which the first seems most popular. However, it has already been observed that simpler and more comfortable setups allow estimation of those parameters. Therefore, there is an increasing spectrum of methods and devices available to perform a HST.

The American Academy of Sleep Medicine suggested in its guidelines four types of devices: in-laboratory, technician-attended, overnight PSG (Type I); full PSG outside of the laboratory not needing a technologist's presence (Type II); devices not recording the signals needed to determine sleep stages or sleep disruption, typically including respiratory movement and airflow, heart rate or ECG, and arterial oxygen saturation (Type III); and those recording 1-2 variables and without a technician, typically arterial oxygen saturation and airflow (Type IV) \cite{AASM}. Therefore, for Types III and IV, there are many novel applications, sometimes even failing to be accurately classified, e.g., peripheral arterial tonometry \cite{PAT}, or audio-based technologies \cite{Glos}.

Even if using these techniques, one can measure sleep for several nights and calculate sophisticated parameters that assess the statistics over many nights. The AHI values are the starting point, often only as raw values, and not connected to clinical ranges. 

It should be mentioned, that there are also studies for which the statistical analysis is reported in the correct, more specific, manner. E.g.,\textit{Yuceege et al.} reported results of qualitative analysis, such as sensitivity, specificity, PPV or NPV \cite{Right}. This is consistent with the newest methodological recommendations presented by \textit{Miller et al.}, who stated that correlational analyses should be conducted alongside qualitative analysis during validity testing \cite{Recommend}.

In addition to that, we proposed a list of possible parameters and approaches, consisting of well known, modified, or heuristically deduced parameters, which can be considered by physicians and statisticians, particularly for comparing and validating various techniques and methods. 

We can state general interpretations of the proposed parameters to allow non-statisticians understanding the course of results. When analyzing the intercept of the linear model - the closer to zero, the better. For the slope of the linear model with a zero intercept and for all correlation analyses (Pearson's, Spearman's or Lin's), or for the bias correction factor - the closer to one, the better. Next, a T/Wilcoxon test p-value greater than 0.05 indicates no reasons to reject the null hypothesis that two means or medians are equal.

The mean value of the differences calculated during Bland-Altman analysis should be as close to zero as possible; however, one should also assess the distribution of the points vs. the mean of all pairs (there should be no relation). The spread of the differences in the Bland-Altman plot should be as low as possible, probably not greater than 20 (it is related with the number of subjects, that participated in the study). Then, the slope in the modified Bland-Altman plot should be as close to zero as possible. 

The simple heuristic ratio should be close to one, which means the distribution of points below and above the Y=X line is similar. The smaller the MAE/eMAE, the better (but the values should be primarily used in relation to other studies). In our opinion, cubic type of ranking function is the most restrictive in terms of its shape (hence chosen as a default one), and sinusoidal - the most liberal, excluding regular MAE. Finally, the higher the accuracy, sensitivity, specificity, Cohen’s Kappa or multi-class AUC, the better.

We do not mean to endorse a single parameter over the range of options that allows considering different contexts of the analysis. Of course, all quantitative approaches remain sensible when considered only within specific subranges of AHI values; however, we omitted such analysis in order to preserve clarity and readability.

It is also important to remember that, in this type of analysis, outliers can have a very large impact. It is possible to estimate the distribution of the "lo-factors" (coefficients indirectly assessing the distance between data points) and to choose the cut-off threshold to remove observations with higher lo-factor values \cite{LOF}. However, as caution in using this option is recommended, we decided not to make it available in this version of the Shiny web app \cite{Shiny}.

Also, the American Academy of Statistics has proposed that the Bayesian approach be used in similar research \cite{Bayes}. However, based on the presented distributions of AHI points in the studied populations, we think that adopting an appropriate prior distribution for such analysis could be difficult.

Gathering all, we decided to introduce a ranking function that can focus more on those results that may ultimately lead to incorrect diagnoses, not only numerical and statistical differences. Surely, it would be also interesting to check whether there is an effect of age and gender on the results. We are considering introducing such functionality into Shiny applications in the future. For now, this can be done "manually" by appropriate division of the input data.

We would also like to emphasize, that the primary aim of this paper was not to compare studies, from which we took the data (noting that they were performed using different methods, on different study groups, and with slightly different motivations). Rather, we aimed to show the application that can evaluate a dataset "statistically" and "clinically" (and compare it with a reference study) in general. Chosen two datasets are only examples, taken for illustration purposes. 

It should also be added that the presented consideration may be extended to different areas of research, where comparative analysis is part of the process.

\section{Summary}

This paper presents the ways in which AHI parameters established by new devices and methods can be compared with the gold standard, reference method. We tried to pay close attention to clinical significance of the AHI results, that are compared; various methods were collected together with a discussion of the interpretation of their results. We also propose a ranking function and modified Bland-Altman plot to extend the analysis. In order to speed up the analysis process, the Shiny web application was prepared for both clinicians and data scientists. 

\section*{Acknowledgment}

The work is part of a short-term scholarship project (The Bekker Programme) funded by the Polish National Agency for Academic Exchange (NAWA). The authors thank Martin Berka for linguistic adjustments. We would also like to thank Dr. Hiroshi Nakano, Dr. Tomokazu Furukawa, and Dr. Takeshi Tanigawa for sharing their data for the presentation.

\section*{CRediT Author Statement}
\textbf{Marcel Młyńczak:} Conceptualization, Methodology, Software, Validation, Formal analysis, Investigation, Resources, Data curation, Writing – original draft, Visualization, Project administration;  
\textbf{Tulio A. Valdez:} Conceptualization, Validation, Resources, Writing – Review \& Editing, Project administration; \textbf{Wojciech Kukwa:} Conceptualization, Methodology, Validation, Investigation, Data curation, Writing – original draft, Supervision, Project administration

\lhead{}
\rhead{}

\end{document}